\documentclass[10pt]{article}
\usepackage{epsfig}
\usepackage{graphicx}
\title {Long term time variability of cosmic rays and possible relevance to the 
development of life on Earth}
\author {A.D.Erlykin $^{1,1a}$ and A. W. Wolfendale $^{1}$\\
$(1)$ Department of Physics, Durham University, Durham, UK\\
$(1a)$ Permanent address: P N Lebedev Institute, Moscow, Russia}
\date{\today}
\begin{document}
\maketitle
\footnote{Corresponding author: Erlykin A.D., e-mail address: erlykin@sci.lebedev.ru, 
telephone: +7-499-135-87-37, fax: +7-499-135-78-80}
\begin{abstract}
An analysis is made of the manner in which the cosmic ray intensity at Earth has varied
 over its existence and its possible relevance to both the origin and the evolution of 
life.

Much of the analysis relates to the 'high energy' cosmic rays ($E>10^{14}eV;=0.1PeV$) 
and their variability due to the changing proximity of the solar system to supernova 
remnants which are 
generally believed to be responsible for most cosmic rays up to PeV energies. It is 
pointed out that, on a statistical basis, there will have been considerable variations 
in the likely 100 My between the Earth's biosphere reaching reasonable stability and 
the onset of very elementary life. Interestingly, there is the increasingly strong 
possibility that PeV cosmic rays are responsible for the initiation of terrestrial 
lightning strokes and the possibility arises of considerable increases in 
the frequency of lightnings and thereby the formation of 
some of the complex molecules which are the 'building blocks of life'.

Attention is also given to the well known generation of the oxides of nitrogen by 
lightning strokes which are poisonous to animal life but helpful to plant growth; 
here, too, the violent swings of cosmic ray intensities may have had relevance to 
evolutionary changes.

A particular variant of the cosmic ray acceleration model, put 
forward by us, predicts an increase in lightning rate in the past and this has been 
sought in Korean historical records.

Finally, the time dependence of the overall cosmic ray intensity, which manifests 
itself mainly at sub-10 GeV energies, has been examined. The relevance of cosmic rays 
to the 'global electrical circuit' points to the importance of this concept.
\end{abstract}


\section{Introduction}
\subsection{The cosmic radiation and its relation to climate}
Over the 4.5By since the formation of the Earth the astronomical environment has been 
variable and, with it, the cosmic ray (CR) spectrum (~this is in addition to solar 
irradiation changes caused by the varying Sun-Earth distance and Earth obliquity -
the 'Milankovich effect'~). There are three main sources of CR variability: the 
Geomagnetic field, the Sun (~by way of the solar wind and the occasional 'solar cosmic
rays' associated with solar flares~) and the presence of nearby CR sources.  
The first two relate to variations of the low energy particles, principally below 10 
GeV, and the last-mentioned to all energies, with increasing variability as the energy 
increases, the highest energy recorded being $\sim 10^{20}$eV (~ie $10^{11}$GeV~).

Further remarks on 'cosmic rays' are necessary. Under 'normal circumstances' 
(~quiescent Sun~) the CR, composed mainly of hydrogen, helium and heavier nuclei, have 
a power law spectrum from about 1000 GeV to 3 PeV. At energies above 3 PeV the spectrum
 steepens. Below 1000 GeV there is a progressive 'flattening' of the spectrum (~as one 
proceeds
towards lower energies~) due to both the Geomagnetic field and the solar wind. The 
magnitude of the flattening depends on the 11 year cycle of the solar wind and is a 
function of geographic latitude and energy; it is very small above 100 GeV.

The energy content of Galactic cosmic rays (~GCR~) is an important parameter. The 
energy densities at earth above the energies indicated are, in units of $Jm^{-3}$, 
 $\sim 0.05 (>10^8eV), \sim 0.03 (>10^{10}eV), 3\cdot 10^{-3} (>10^{12}eV), 10^{-4} 
(>10^{14}eV)$ and $10^{-6} (>10^{16}eV)$ (Wolfendale, 1973). By contrast, the total 
solar irradiance is some $10^8$ times greater than that for all CR.

In addition to an interest in its own right, the variation of the GCR energy spectrum 
with time has relevance to atmospheric properties and thus (perhaps) to `life' on 
Earth.  The first mention of a possible effect of CR on climate seems to be 
that of Ney (1959), the claimed mechanism being by way of the effect of CR ions on 
aerosols, leading to enhanced cloud formation. A number of workers have, more recently,
 followed up the suggestion (eg Svensmark and Friis-Christensen, 1997; Palle Bago and 
Butler, 2000; Marsh and Svensmark, 2000; Svensmark, 2007). Indeed, Svensmark (2007) has
even coined a term for the new discipline: 'Cosmoclimatology' !

Although, a priori, it might be thought that such a mechanism was unlikely, based on 
the $10^8$ factor referred to above, Voiculescu et al.(2006) have claimed that there is
 a GCR, low cloud cover correlation over restricted regions of the Globe. Harrison and 
Ambaum (2008) have claimed that 'the mechanism' exists on the edges of clouds by way of
 a large reduction in the critical supersaturation needed because of a large degree of 
droplet charging. 

This is where it must be pointed out that direct CR may affect the climate. The 
effect of GCR on the 'Global electrical circuit' (~Williams, 2002; Rycroft et al., 
2008~) has been studied by Tinsley (2008) and Aplin et al. (2008) who consider the 
effect of 'electro-freezing' and 'electro-scavenging'; these processes lead to changes 
in the density of cloud condensation nuclei. In all these processes it is important to know the manner in which CR ionization varies with atmospheric depth. Such studies have been made by Usoskin and Kovaltsov (2006) and Velinov et al.(2009).

Despite doubts about the appreciable effect of GCR on clouds in the lower troposphere,
(~Sloan and Wolfendale, 2008; Erlykin et al., 2009b~), in the stratosphere, where GCR 
intensities are higher and there are the occasional 'solar protons', there are strong 
signals. That
there is an 11-year cycle in the ozone density is beyond doubt, and Lu (2009) has 
presented strong evidence for a strong causal correlation between GCR and polar ozone 
loss over Antarctica (~the 'ozone hole'~). The identification of GCR as being 
responsible, as distinct from solar irradiance, comes from CR being the only source of 
low energy electrons at the depth in question and the ozone hole being in the 
Geomagnetic Pole region where the CR intensity is a maximum. Other related effects 
include the effect of strong Geomagnetic storms and Forbush decreases of GCR intensity 
on the total ozone content and the lower atmosphere (~troposphere and lower 
stratosphere~) (~Lastovicka and Krizan, 2005~). It must be mentioned, however, that the
 claimed effect of Forbush CR decreases on the liquid cloud fraction in the troposphere
 (~and other atmospheric parameters~) by Svensmark et al. (2009) was not confirmed by 
Laken et al. (2009).

A less certain, but potentially important, process, is that suggested by Shumilov et 
al. (1996). These workers were impressed by the increase in aerosol concentration after
 solar proton events (~particularly the 'Ground Level Event' of $16/02/1984$. The 
increase occured at the altitude range, $\sim$17km, where energetic solar protons lose 
their energy in the atmosphere~). The mechanism put forward was CR ionization, as a 
source of ion nucleation, stratospheric sulphate aerosols forming on the condensation 
nuclei (~Arnold, 1982; Hofman and Rosen, 1983~).

The importance of solar particle events in polar regions has been pointed out by 
Usoskin et al. (2009). A related argument, which may have relevance to the claimed low 
cloud cover, GCR correlation, is due to Kudryavtsev and Jungner (2005). These workers 
argue that the extra CR induced aerosols cause atmospheric transparency changes which 
in turn affect tropospheric climate. They quote other workers (~eg Starkov and 
Roldugin, 1994 and Pudovkin et al., 1997) as having also observed transparency 
decreases during solar proton events.
\subsection{Extensive Air Showers}
Our work reported here relates mainly to much higher energies than those concerned with
 solar effects (~which are mainly below some 10s of GeV~), specifically 0.1PeV and 
above; these particles being manifest by their production of extensive air showers 
(EAS). Some remarks about EAS are necessary. When a particle 
(~often a proton~) of 'high energy', say above 0.1PeV, is incident on the atmosphere it
 interacts with the air nuclei to produce a cascade of secondary particles (~mainly 
pions of the three charge states: positive, negative and zero~). The (unstable) charged
 pions decay into muons ('heavy electrons'), which in turn may decay into electrons,
 and the neutral pions decay into gamma rays. The process is repeated by the primary 
proton which survives the interaction with reduced energy and those energetic pions
 which interact further before they have had chance to decay. 

The effect of the interactions is to build up an 'electromagnetic cascade' of electrons
 and gamma rays. To be quantitative, a 1PeV primary will generate a shower having 
maximum number of charged particles (~mainly electrons~) at an atmospheric depth of 
$\sim$560mb, ie height in the atmosphere of $\sim$5km. The mean number of particles at 
'shower maximum' would be $\sim 5.3\cdot 10^5$ and the number of particles at ground 
level for a vertically incident proton would be $\sim 1.4\cdot 10^5$ (Ambrosio et al.,
1997).

The important feature of EAS relevant to the initiation of lightning strokes is the 
high density of secondary electrons near the axis of the shower. Recent calculation by 
us (~Erlykin et al.,2009a~) for 100PeV protons and a height of 2km give a particle 
density at 1m from the shower axis of $\sim 2\cdot 10^5 m^{-2}$. For a primary of 
energy 1PeV the density will be $\sim 2\cdot 10^3 m^{-2}$, still a very high value. On 
the axis itself, the particle density will be some ten times greater.

\subsection{Relevance of EAS to the origin of life}
The possible relevance to the atmosphere (~and `life', including humans~), is 
by way of the very likely role of EAS particles in the initiation of lightning (eg 
Gurevich and Zybin, 2001, Chubenko et al, 2009, Gurevich et al, 2009 and Chilingarian 
et al, 2009).  The idea is that the leader lightning stroke is initiated by runaway 
electrons which are generated by particles in EAS.  The references quoted include 
observed coincidences between EAS and lightning and not just the undoubted effect of 
thunderstorm electric fields on the energies of CR particles (~which are, themselves, 
not necessarily members of EAS~). The whole question of the electrical conditions of 
the atmosphere, including its most dramatic manifestation (~lightning~), is tied up 
with CR insofar as they represent an important source of ions near ground level and the
 major source at altitudes above a few km. Tinsley et al, (2007), Rycroft et al, 
(2008), and others have pointed out the great importance of the `global electric 
circuit' - to which CR contribute considerably - even when the changes considered have 
been small. The global electric circuit possibly can be influenced also by gigantic red
 sprites and blue jets. Presumably they are initiated by electrons, which are created 
by CR and accelerated in their upward movement to runaway energies by thunderstorm 
electric fields (~Yukhimuk et al. 1998; Tonev and Velinov, 2003~). Effects consequent 
upon very large changes in CR intensity at Earth could be profound. The question of the
 global electrical curcuit is considered in more detail later.

Lightning has, conceivably, played a role in the evolution of life.  Starting with 
pre-life, the work of Miller and Urey (~Miller, 1953~) involving the passage of 
electrical discharges through a  `pre-biotic soup' of appropriate chemicals (~water, 
methane, ammonia, etc~) caused quite complex molecules to be generated: amino acids, 
monomers, RNA etc, which were necessary pre-cursors of elementary life. Lightning 
could, conceivably, have provided the required discharges. It must be remarked that 
there are different views about the origin of life; some argue that the initial complex
 molecules arrived by way of comets instead (~eg Hoyle and Wickrama-singhe, 1993~).  
Here, we persist with the Miller and Urey hypothesis (MU) since very recent work (~eg 
Parman, 2009~) concludes that there was little free oxygen in the atmosphere prior to 
2.45 Gy BP (~ie $2.45\cdot 10^9$ years before present~) and the MU hypothesis would 
have a chance of success.  The necessary water oceans were probably present by 4.2 Gy 
BP (`Bada, 2003; Parman, 2009~).

Later, when `life' was advanced, lightning would have had an effect on evolution by 
virtue of the obnoxious NO$_{x}$ (`NO and NO$_{2}$) produced.  Even now, some 20\% of 
NO$_{x}$ comes from lightning - much higher lightning rates would have been important. 
 NO$_{x}$ effects include modifications to atmospheric chemistry, with particular 
relevance to ozone levels and effects on hydroxyl (OH) radicals - thereby increasing 
the concentration of greenhouse gases.

There is a wealth of literature on NO$_x$ production by lightning (~eg Betz et al., 
2008~). Allen et al. (2009) estimate that the rate of production
 of NO$_x$ from lightning is $\sim 10^{13}kg \cdot year^{-1}$, to be compared with the 
total atmospheric mass of $5\cdot10^{18}kg$ and a mass of ozone of order 
$5\cdot10^{11}kg$ (~Allen, 1973~).

Although NO$_x$ is damaging to mammals, plants benefit from the nitrates coming from 
NO$_{x}$ reactions. The interplay between the development of plants and of mammals 
means that periods of low CR intensity (~low NO$_{x}$~) as well as those of high 
intensity are important. The likelihood of CR effects here is, no doubt, less 
contentious and should be put alongside the various meteorological factors, such as 
temperature and rainfall, which have affected the evolution of life.

Even if none of the above effects turn out to be important, a knowledge of the past 
history of the intensity of high energy GCR (HECR), by which we mean $10^{14}$ eV and 
above, is of considerable interest because of its relevance to the (~still unsolved~) 
problem of the origin sites, acceleration mode and propagation characteristics of the 
primary particles.
\subsection{Scope of the paper}
We start with an analysis of the time variation on a statistical basis using results 
provided by us earlier (~Erlykin and Wolfendale, 2001a~), and based on our supernova 
remnant (SNR) model of GCR acceleration (~Erlykin and Wolfendale, 2001b~). Later we 
examine the recent past - some 30,000 y - assuming that our Single Source Model of the
 `knee' in the spectrum at $\sim$3PeV (~Erlykin and Wolfendale, 1997, 2003~) is 
correct. In this model,
 which is now being increasingly accepted (~eg Hu, 2009~), we argue that the extreme 
sharpness of the transition region of the energy spectrum is indicative of the presence
 of a recent, nearby SNR.

Finally, some remarks will be made about the possibility of the total intensity of CR, 
as distinct from just the high energy component, having relevance to the terrestrial 
climate and thereby to evolutionary mechanisms. This is left to last because we are 
less convinced by the claims for its relevance to the lightning hypothesis but it is 
included for its relevance to the other mechanisms.

\section{Variations of Galactic cosmic rays over the past million years using the results of Erlykin and Wolfendale (2001a)}
\subsection{Time profiles as a function of energy}
It is assumed at this stage that the Geomagnetic field is constant and that the solar 
wind modulation is `normal', viz giving only an 11-year variation in total CR intensity
 with peak-to-peak magnitude of, typically, 10\%.  Small, long term variations of solar
 irradiance are ignored.  The CR variations are thus due to the changes in the Galactic
 component.
 
 Figure 1 shows a typical time-profile of the GCR intensity for different energies from
 our supernova model for CR production, Erlykin and Wolfendale (2001a, b).  Protons are
 assumed in the calculations but the results can be applied to other nuclei by simple 
rigidity-transformation (~rigidity$=E/Z, E$ and $Z$ being particle energy and charge 
respectively~). It will be noted that, in addition to the 
rare upward excursions, which are particularly marked at the highest energy (~taken 
here as 10 PeV, ie $10^{16}$eV~), there are long periods - by chance - when the average
 level is well below the long term average value. The physics behind the intensity 
behaviour shown in Figure 1 is straightforward and will be described.  It is well 
known that the diffusion coefficient of GCR varies with energy, increasing as the 
energy increases.  Thus, the `wave' of GCR from a source (~SNR in this case~) 
propagates more rapidly, and with a narrower time width, at high energy.  This is 
apparent in the Figure, where the high energy `spikes' are higher and narrower than the
 low energy ones.  It should be added that although we have taken the sources as being 
SNR - the most likely situation - a similar type of source, such as a pulsar, would 
give a similar result.

The frequency of excursions in CR intensity can be examined as follows. Starting with 
the positive excursions, we define `peaks' above nearby minima and give a $logN, logS$
 plot, where $S$ is the intensity of a peak and $N$ is the number of times such a peak 
intensity, or bigger, is achieved.  The result is shown in Figure 2.

We remember that the data are binned in 1000y (~ie they represent the average intensity
 over such periods~).  The lines drawn in Figure 2 are simple
 parabolic `best-fits'. An indication that the calculations are correct comes from an 
examination of the slopes of the tangent in Figure 2, ie the $\gamma$-values, where 
$N(>S) \propto S^{-\gamma}$, ie $logN(>S)=-\gamma logS + constant$. In the `middle 
region', say $log S$ = 1 or 2, the shape should follow a line of slope 1 since, here, 
we are dealing with, essentially, a two-dimensional distribution of sources (SNR), 
these being mainly further away than the half-thickness of the SNR distribution about 
the Galactic Plane (~half width at half maximum $\simeq 250$ pc~). The reason is 
straightforward: a 
source at distance $x$ will give $S(x) \propto 1/x^2$ so that sources within $x$, of 
number $\propto x^2$, will have intensity $>S$, thus $N(>S) \propto x^2$ ie 
$\propto 1/S$; $\gamma=1$. At $S$ values below 1 the curvature arises from the loss of 
small $S$-values due to `source-confusion'.  Eventually, above $logS$ = 2, the slope 
should tend to -1.5 because some of the sources will be nearer than 250pc and the 
distribution of relevant sources tends to isotropy (~the argument is similar to that 
for 1-D with the number of sources within distance $x$ being $\propto x^3$~).

Of particular interest is the extension to cover a period of 100My, the likely `window'
 when other conditions on Earth were suitable for elementary life to form.  Presumably 
(but not definitely) this was immediately after the `late heavy bombardment' some 3.9Gy
 before present (~Parman, 2009~). It can be remarked that the earliest fossils date 
from about 3.5Gy before present.  We note than in 100My of order one peak would occur 
for energy above 10PeV with intensity some 3000 times the datum. Taking the median 
value of $logI$=1.69, the enhancement is a factor of about 60. Such an enhanced 
intensity would continue for a few thousand years.

\subsection{Temporal effects}
It is of relevance to examine the fraction of time for which the CR intensity would be 
above and below certain limits, over our `standard' period of 1My.  This is given in 
Figure 3 for particle of energy 10PeV.  It will be noted that for 10\% of the time the 
intensity will be more than ten times the median and for 10\% of the time, the 
intensity would be less than one quarter of the median.

\subsection{Short-term variations for 10,000 year bins}
Figure 4 shows the equivalent to Figure 1 for time bins which are ten times that used 
previously, viz now 10,000 years.  The 10PeV peaks are typically 5 times smaller than 
in Figure 1.  The equivalent of Figure 2 would give an enhancement by about a factor of
 at least 10, lasting 10,000y every 100My.

\section{The likely high energy GCR intensity in the immediate past}
In earlier work (~Erlykin and Wolfendale, 2003~), we identified the `single source', 
responsible for the `knee' in the cosmic ray spectrum as probably being  a supernova in
 the distance range 250 - 400 pc from the Earth and being of age in the range 85 to 115
 ky (~1pc = $3\cdot 10^{16}$m~). Figure 5 shows the results of our calculations for a 
distance of 300pc and the range of ages just indicated. It will be noted that the 
ratio of the predicted intensity (~for 10PeV~) at the peak to that at present covers a 
wide range: from 10 to 1000.  Certainly, in the `recent past' (~some thousands to tens 
of thousands of years~), the intensity of high energy GCR should have been 
significantly higher than at present.

Studies have been made using radioactive nuclei in ice cores of different ages of past 
CR intensities but these refer to low energy particles. At this stage of the present 
work, with the emphasis on high energy particles ($\simeq$ PeV), we have examined 
historical records over the past centuries of both lightning frequency and other, 
relevant, atmospheric phenomena. Classical literature is replete with mention of 
lightning and other dramatic atmospheric phenomena. Zeus was the Athenian God of 
lightning and, for example, Chaak was the Mayan Lightning God (~Looper, 2003~), the 
period of relevance being the 8th Century AD. However, apart from one source of 
information, the records cannot be used to give lightning rates (~Stephenson, private 
communication, 2009~).
The exception relates to data from Eastern chronicles, which date from about 1400, 
specifically, the annals of the Chosun-Dynasty relating to the Korean Peninsula are 
comprehensive and have been analysed by Lim and Shim (2002); these authors remark that 
the 'time variation of the (climate) indices shows a good agreement with a similar 
analysis done for China by other authors'. There appears to be nothing of significance 
for other areas of the Globe (~Stephenson, 2009, private communication~). 

Returning to the Korean records, Lim and Shim (2002) have given such 
information for the period 1400 - 1900 and this shows interesting variability. It is 
true that the average rate of lightning per century was higher in the past but the 
variability was so dramatic as to make any quantitative estimate of the likely 
CR-induced variation suspect.  For example, with respect to the `final' period, 1800 - 
1870, where the mean rate was given as about 1 stroke per year, the mean values for the
 factors of increase were as follows:
\begin{itemize}
\item 1400 - 1500: 3 (3)
\item 1500 - 1600: 7 (5) (with occasional yearly frequencies above 20 strokes per year)
\item 1600 - 1700: 4 (2)
\item 1700 - 1800: 5 (3)
\end{itemize}

No doubt, various meteorological factors unrelated to cosmic rays were responsible for 
at least some of the variability and we have endeavoured to take out some of this 
variability by dividing the lightning frequency by the annual occurence of rain and 
snow from the same source of data. The ratios, again with respect to the 1800 to 1870 
period, are shown in brackets. It will be noted that the two sets of figures are rather
 close. 

Contemporary values of lightning frequency would allow a check on the possibility of a 
'time-gradient' in lightning frequency but these are not available. Furthermore, there 
are difficulties in ensuring consistent criteria on what to include in the 470 year 
long record (Jongman Yang, 2009, private communication). Although the fact that China 
seems to have shared the
variability leads us to believe that the mean lightning rate on a Global scale was 
probably somewhat higher than recently. It must be said, however, that there can be no 
question (~yet~) of defining the distance and age of the single source, using this 
method. More extensive studies remain to be done. 

\section{Variation of the intensity of low energy cosmic rays}

\subsection{General Remarks}
As remarked earlier (~Section 1~), low energy CR are modified in intensity by the 
Geomagnetic field and the solar wind as well as, to a lesser extent, by nearby SNR.  
To this should be added solar CR themselves.  In all cases there may be 
relevance to the main thrust of the present work as will be demonstrated.

\subsection{The Dwyer-model for lightning activity}
Dwyer (2005) has claimed that non-uniformities in the atmospheric electric field can be
 amplified by the `steady background of atmospheric cosmic rays' and thus influence 
lightning activity. Satori et al.(2007) have, indeed, found evidence for annual and 
semiannual areal variations of Global lightning on an 11-year cycle although we, 
ourselves, have failed to find the expected CR - correlated variation of lightning 
frequencies over the geographical land masses. It must be said, however, that these 
have been reports, by Stozhkov (2003), for an increase in the GCR intensity 
leading to a growth of thunderclouds.

What is clear is that the CR intensity (~Galactic or Solar~) will have an effect on the
 Global electric circuit, as already mentioned in connection with the effect on the 
charging of condensation nuclei. The effect will be on both the 'fair weather field', 
by virtue of CR ionization (~Kniveton et al., 2008~), on the field in clouds through the changes to clouds 
referred to already and to the 'current generators' in thunderclouds (~Tinsley et al., 
2007~) if the Dwyer model is, indeed, applicable.

If there is indeed a dependence of lightning frequency on the total CR intensity, as 
distinct from on high energy GCR intensity, then interest will focus on variations of 
the Geomagnetic field (reversals) and of the solar wind - both of which can be 
considerable. However, it should be noted that the effects here will be confined 
largely to high latitudes and the higher regions of the atmosphere where low energy CR 
are involved.

\subsection{The role of magnetic field reversals}
The field reversal, as such, has no effect on the surface level CR intensity in the 
sub-100 GeV region, rather it is the period between reversals when the field is at a 
low level. A change in CR intensity by a factor less than about 3 would be expected.  
`Contemporary' reversals number about 200 over the period of about 165My for which data
 are available (Creer and Pal, 1989).  Such variations in CR rate are probably not of 
great importance from the evolutionary point of view. Of greater relevance would be the
 (~likely~) dramatic changes in the Geomagnetic field during the early stages of the 
Earth's formation, but these are of unknown magnitude.  What can be said is that they 
would presumably lead to large reductions in the CR rate.

\subsection{The solar wind}
Modifications to the solar wind by intrinsic solar changes and changes to the 
interstellar medium in which the solar system is immersed will cause GCR changes.  The 
former, which may give rise to solar flares, will be considered in the next section. 
Concerning the latter, Vahia (2006) has made a detailed study of the expected 
modification to the 
GCR spectrum at Earth as the solar system passed through various environments : spiral 
arm, interarm, a Giant Molecular cloud and the immediate vicinity. Over the last 10My
the local interstellar medium density has varied from 0.08cm$^{-3}$ in the 'Local 
Fluff' (~Frisch, 1995~), the weak local interstellar cloud which provides the pressure 
for the confining solar wind bubble and occupies about 5pc, through 4cm$^{-3}$ for the 
Geminga SNR to $5\cdot 10^{-4}$cm$^{-3}$ for the interarm region. The result is that 
above 100 GeV there is no change but at an energy 
of 3 GV, typical of the `total CR intensity', the reductions in intensity cover a range
 of about 10.  At 300MeV, an energy of relevance to particles entering in Polar 
regions, the range is even bigger and is about 100.

The times taken to cross some of the regions listed above cover the range 0.1 - 20My so
 that for such periods the CR intensity will have fallen by up to one or two orders of 
magnitude.

Concerning the spiral arm$/$interarm regions, mention should be made of the works of 
Shaviv (2002, 2003) and later publications. The argument related to
 the periodic crossings of spiral arms where the GCR intensity is higher than in the 
interarm regions. It was claimed that the 'icehouse' episodes, which numbered 4 in 
500My, were due to the enhanced GCR intensity causing more low cloud and thus lower 
ground level temperatures. This argument has been elaborated upon by Svensmark (2007) 
in his 'Cosmoclimatology' works. However, there are problems. Firstly, our own work 
(~unpublished~) using cosmic gamma ray data from the EGRET instrument on the Gamma Ray 
Observatory (~Hunter et al., 1997~) shows that enhancement in the sub-10GeV region is 
less than a factor 2. This result is consistent with our earlier work (~Rogers et al., 
1988~). Interestingly, there should be a bigger arm, interarm contrast in the PeV 
region because of the higher density of supernovae in the spiral arms but the rapid 
spatial diffusion of these particles reduces its magnitude. Secondly, Melott et al. 
(2009) have used recent CO data (~which provides information about molecular hydrogen 
in the Galaxy and the densities of young stars~) to make more 
robust estimates of the positions of the spiral arms and the times of transit of the 
solar system; these estimates cause the claimed correlations of arm transit and 
icehouse estimates to disappear. Melott et al. argue that 'the correlations cannot be 
resurrected by any reasonable pattern speed'.

\subsection{Solar Flares}
Very large flares can, in principle, cause big changes to the low energy CR intensity. 
 A number of workers have considered the topic of `cosmic rays and ancient 
catastrophes'.  Wdowczyk and Wolfendale (1977) drew attention to the fact that the 
`$log N, log S$' curve for solar CR, where $S$ is the `fluence' (~ energy per unit 
area~), is linear, at Earth, up to the strongest flare recorded in the period from 
1956. The 
maximum value is $1 Jm^{-2}$, averaged over the Earth's surface, and the corresponding 
rate is about $2\cdot 10^{-2}y^{-1}$.  The current ambient CR energy intensity is 
$1 Wm^{-2}$ to that, for a period of 10h (a typical flare length), the fluence is 
$4\cdot 10^{-2} Jm^{-2}$. The strongest flare so far recorded therefore corresponds to 
a 25-fold increase in CR intensity (~at the Earth's surface~) for this period of 10h.  
Extrapolation to the inevitable stronger flares with much lower frequency is impossible
 with any accuracy in view of lack of knowledge of the details of flare acceleration. 
However, extrapolation using an exponential fall beyond the maximum flare fluence seen 
so far would indicate a fluence of $10^5 Jm^{-2}$ every 100My on average. Such a 
fluence would correspond to a radiation level of $\sim 10$ R\"{o}ntgen, a serious dose 
for mammals. The derived rate would not be inconsistent with the results of $^{10}$Be 
and $^{26}$Al studies in sea sediments.
\subsection{Very local supernovae}
Wdowczyk and Wolfendale (1977) and others have derived the fluences of CR at earth, for
 both particles and gamma rays, which would result from close proximity to a SN.

There will be a weak gamma ray flash, weak because of the absorption close to the 
source, the average fluence for a SN which has 'unit probability' of being seen in 
$\sim$100My being $\sim 10^3 Jm^{-2}$. Insofar as the 'flash' lasts for several hundred
 days the contribution would be small in comparison with the ambient CR intensity of 
$\sim 1Wm^{-2}$, ie $\sim 10^5 Jm^{-2}$ over 100 days.

For particles, the changes in CR intensity are as given in the Figures, ie the spikes, 
which have magnitudes and time widths which are a function of particle energy. 
\subsection{Gamma Ray Bursts}
Concluding this brief discussion of sub-10GeV CR and their possible effect on the 
Earth's atmosphere and the Earth's 'inhabitants', mention must be made of 'gamma ray 
bursts'. Although most bursts are at cosmological distances it is possible that there 
has been one or more bursts from the Galactic Centre, the most recent having been 
$\sim$12My ago (~Sanders and Prendergast, 1974~). Wdowczyk and Wolfendale (1977) have 
examined this possibility and derived a fluence of 1MJm$^{-2}$ at the top of the 
atmosphere. The corresponding radiological dose is $\sim$100R. Although it is true that
 the gamma rays, being in the MeV region, will be largely absorbed by the 20-30km 
altitude region their effect on the ozone layer and the dynamics of the atmosphere 
(~see Section 5~) make an effect on the troposphere inevitable. There is much 
information about CR effects on lunar rock, principally from solar CR bursts, but there
 are, as yet, no definitive records of gamma ray bursts - indeed one could visualise 
determining the energy spectrum of CR from the cascade damage in the lunar regolith 
region if quantitative measurements were possible. It is hoped that this situation will
 be realised.

\section{Discussion and Conclusions}
The two energy ranges can be considered in turn: the PeV region, where the occasional 
excesses and deficits of GCR, are relevant, and the sub-10 GeV region where 
solar-induced phenomena are important.

In both cases we are concerned mainly with induced lightning, the PeV region relating 
to the Gurevich effect of EAS cores initiating lightning and the sub-10 GeV region 
relating to the claim by Dwyer and others that the overall-CR-induced ionization level 
is important for lightning generation.

In neither case is it evident that the lightning rate would be proportional to the CR 
intensity. Although detailed calculations for the Global electric circuit have been 
made (~the 'EGATEC-model' of Odzimek et al.,2009~) the functional form has not yet been
 derived (~Odzimek, 2009, private communication~). However, it is evident from the 
physics of the lightning process that many regions with thunderclouds which were not 
hosts to lightning would become so if the PeV GCR rate were to increase. A similar 
situation would be expected for the sub-10GeV case.

Starting with the PeV region, the results are, from the GCR point of view, 
straightforward: considerable fluctuations in PeV GCR intensities should occur over 
long periods of time (My). At high energies, in fact, the variations would be bigger 
than quoted if, as seems possible, the diffusion coefficient in the `local bubble' in 
the interstellar medium, in which we reside, were higher than the conventional one - 
for a uniform interstellar medium - adopted in our calculations.

It can be remarked that, since most of the fluctuations are stochastic and geometrical 
in origin, CR production by other types of `discrete' sources, such as pulsars, would 
give rather similar results. The very close SNR responsible for the dramatic upward 
high energy CR intensity fluctuatiions are unlikely to have given dramatic `gamma ray 
flashes' which could have had an effect on the Earth. A gamma ray burst at the centre
 of the Galaxy - for which there is no direct evidence - could have been 'serious', 
however.

Turning to the relevance of the results to lightning and to possible biological effects
, in the 100My window for life creation, the considerable increase in 10PeV intensity
 for some tens of thousand years, with its presumed increased lightning rates - could 
have played a part in pre-biotic life generation.

At later stages, when life was evolving, the occasional lightning excesses with 
increased production of NO$_{x}$ could have had pronounced positive effects on 
vegetation and negative effects on humans.  However, evolutionary spurts for non-plant 
life may have occurred for those long periods when the 10PeV intensity was low. In this
 connection, because CR induced ionization is related to both effects NO$_x$ and 
lightning, the extension of recent models (~Usoskin and Kovaltsov, 2006; Velinov et 
al.,2009~) to higher energies will be of considerable interest.
 
The claimed PeV increase in the recent past could (`over, say, 5000 years~) 
conceivably be found in historical records of changes in lightning rates more extended 
than those carried out by us so far.

Turning to the variations in the sub-10GeV intensity, if, indeed the lightning rate is
 affected by CR of all energies, then, again, CR-induced evolutionary effects are 
expected. These changes would be expected to be confined to high latitudes and high 
altitudes although still important if the claimed mechanisms of transmitting 
stratospheric changes to the troposphere (~eg Haigh, 1996; Kudryavtsev and Jungner, 
2005~) 
are effective. The main causes envisaged relate to the movement of the solar system 
through different environments in the interstellar medium, geomagnetic field reversals 
and solar flares. Even without lightning, changes to the global electric circuit could 
have generated important climatic effects.

In conclusion, it is argued that CR should be considered alongside other astronomical 
factors, most notably changes in the Earth-Sun distance and the Earth's spin axis 
(~'Milankovich effects'~), in causing effects on the initiation and later 
evolution of life.

{\bf Acknowledgements}

The Physics Department of Durham University is thanked for the provision of excellent 
facilities. Drs A Odzimek and K Aplin and Professor A Chilingarian are thanked for 
helpful comments. The authors are grateful to the Kohn Foundation for supporting this 
work and to Professors Jongman Yang and F.R.Stephenson for helpful advice.

\newpage
{\bf References}

\begin{itemize}
\item[1.] Allen, C W, 'Astrophysical Quantities', Athlone Press (1973).
\item[2.] Allen, D, Pickering, K, Pinder, R and Pierce, T, 'Impact of lightning - NO 
emission on eastern United States photochemistry during the summer of 2004 as determined using the CMAS model', Proc. 2009 CMAS Meeting.
\item[3.] Ambrosio, M, Aramo, C, Colesanti, L, Erlykin, A D and Machavariani, S K,
'Frontier Objects in Astrophysics and Particle Physics', Eds. F. Giovannelli and G. 
Mannocchi, Italian Phys. Soc, 57 (1997), 437.
\item[4.] Aplin, K I, Harrison, R G and Rycroft, M J, 'Planetary Atmospheric 
Electricity', Eds. F.Leblanc et al. doi:10.1007/978-0-387-87664-1\_3
\item[5.] Arnold, F, 'Ion nucleation - a potential source for stratospheric aerosols',
Nature, 299 (1982), 134.
\item[6.] Bada, J,L, `Origins of Life', Oceanography, \textbf{16}, (2003), 3, 98.
\item[7.] Betz, H D, Schumann, U and Laroche, P, (eds) 'Lightning, Principles, Instruments and Applications', (2008), Springer.
\item[8.] Chilingarian, A, Daryan, A, Arakelyan, K, Reymers, A, Melkumyan, L, 
'Thunderstorm correlated enhancements of Cosmic Ray Fluxes detected on Mt. Aragats', 
Proceedings of international conference FORGES 2008, Nor Amberd, Armenia, pp./ 121-126,
 TIGRAN METS, 2009.
\item[9.] Creer, K M and Pal, P C, 'On the frequency of reversals of the Geomagnetic 
Dipole', Catastrophes and Evolution -Astronomical Foundations, ed. Clube, SVM, 
Cambridge University Press, (1989), 113.
\item[10.] Chubenko, A P, Karashtin, A N, Ryabov, V A, Shepetov, A L, Antonova, V P, 
Kryukov, S V, Mitko, G G, Naumov, A S, Pavljuchenko, L V, Ptitsyn, M O, Shalamova, S 
Ya, Shlyugaev, Yu V, Vildanova, L I, Zybin, K P and Gurevich, A V, `Energy Spectrum of 
lightning gamma emission', Phys. Lett. A. (2009) doi:10.1016/j.physleta.2009.06.031.
\item[11.] Dwyer, J R, `The initiation of lightning by runaway air breakdown', Geophys.
Res.Lett., \textbf{32}, (2005) L20808.
\item[12.] Erlykin, A D and Wolfendale, A W, `A single source of cosmic rays in the 
range $10^{15}$ - $ 10^{16}$ eV', J.Phys.G. \textbf{23}, (1997) 979.
\item[13.]  Erlykin, A D and Wolfendale, A W, `Supernova remnants and the origin of the
 cosmic radiation : I SNR acceleration models and their predictions', J.Phys.G., 
\textbf{27}, (2001b) 941.
\item[14.] Erlykin, A D and Wolfendale, A W, `Supernova remnants and the origin of the 
cosmic radiation : II spectral variations in space and time', J.Phys.G. \textbf{27}, 
(2001a) 959.
\item[15.] Erlykin, A D and Wolfendale, A W, `High-energy cosmic gamma rays from the `single-source', J.Phys.G. \textbf{29}, (2003) 709.
\item[16.] Erlykin, A D, Parsons, R D and Wolfendale, A W, 'Possible cosmic ray 
signatures in clouds ?', J. Phys. G: Nucl., Part. Phys., (2009a), 322495/PAP/158830.
\item[17.] Erlykin, A D, Gyalai, G, Kudela, K, Sloan T, and Wolfendale A W, `On the 
correlation between cosmic ray intensity and cloud cover', J.Atmos.Sol-Terr. Phys. 
(2009b), doi:10.1016/j.jastp.2009.06.012.
\item[18.] Frisch, P C, 'Characteristics of nearby Interstellar Matter', Space Science 
Rev., {\bf 72}, (1995) 499.
\item[19.] Gurevich, A V and Zybin, K P, `Runaway breakdown and electric discharges in 
thunder-storms', Physics Uspekhi, \textbf{44}, (2001), 1119.
\item[20.] Gurevich, A V, Karashtin, A N, Ryabov, V A, Chubenko, A P and Shepetov, A L 
(2009), `Non-linear phenomena in the ionsopheric plasma.  Effects of cosmic rays and 
runaway breakdown on thunderstom discharges', Physics Uspekhi, \textbf{52}, (2009) 735.
\item[21.] Haigh, J D, 'The impact of solar variability on climate', Science, 
{\bf 272}, (1996) 981.
\item[22.] Harrison, R G and Ambaum, M P H, 'Enhancement of cloud formation by droplet 
charging', Proc. Roy. Soc. a, doi:10.1098/rcpa 2008.0009.
\item[23.] Hofman, D J and Rosen, J M, 'Condensation nuclei events at 30 km and 
possible influences of cosmic rays', Nature, {\bf 302}, (1983) 511.
\item[24.] Hoyle, F and Wickrama-singhe, N C, `Our place in the Cosmos', pbl.J.M.Dent, 
Phoenix Publ.(1993) ISBN 978 1 861978486.
\item[25.] Hunter, S D, Bertsch, D L, Catelli, J R et al., 'EGRET observations of the 
diffuse gamma ray emission from the Galactic Plane', Astrophys. J., {\bf 481}, (1997)
205.
\item[26.] Hu, H, 'Status of the EAS studies of cosmic rays with energy below 
$10^{16}$eV', (2009), arXiv:0911.3034.
\item[27.] Kniveton, D R, Tinsley, B A, Burns, G B, Bering, E A and Troshichev, O A,
'Variation in global cloud cover and the fair-weather vertical electric field', 
J. Atmos. Solar-Terr. Phys, {\bf 70}, (2008) 1633.
\item[28.] Kudryavtsev, I V and Jungner, H, 'A possible mechanism of the effect of 
cosmic rays on the formation of cloudiness at low altitudes', Geomagnetism and 
Aeronomy, {\bf 45}, (2005) 641.
\item[29.] Laken, B, Wolfendale, A W and Kniveton D, 'Cosmic ray decreases and changes in 
the liquid cloud fraction over the oceans', Geophys. Res. Lett., {\bf 36}, (2009) 
L23803.
\item[30.] Lastovicka, J and Krizan, P, 'Geomagnetic storms, Forbush decreases of 
cosmic rays and total ozone at northern middle higher latitudes', J. Atmos. Solar-Terr.
 Phys. {\bf 67}, (2005) 119.
\item[31.] Lim, Guy-Ho and Shim, Tae-Hyeon, 'The Climate based on the Frequency of 
Meteorological Penomena in the Annals of Chosun-Dynasty', Science in China (B),
\textbf{38}, (2002) 4, 343.
\item[32.] Looper, M G, 'Maya Art and Kingship at Quirigua', Univ. of Texas Press, 
(2003).
\item[33.] Lu, A B, 'Correlations between cosmic rays and ozone depletion', Phys. Rev. 
Lett., {\bf 102}, (2009) 118501.
\item[34.] Marsh, N and Svensmark, H, 'Low cloud properties influenced by cosmic rays',
 Phys. Rev.Lett., {\bf 85}, (2000) 5004.
\item[35.] Melott, A L, Overhott, A C and Pohl, M, 'Testing the link between 
terrestrial climate change and Galactic spiral structure', Astrophys. J., {\bf 705}, 
(2009) L101.
\item[36.] Miller, S L, `A production of amino-acids under possible primitve earth 
conditions', Science \textbf{117}, (1953) 528.
\item[37.] Ney, E P, 'Cosmic Radiation and the Weather', Nature, {\bf 183}, (1959) 451.
\item[38.] Odzimek, A, Loster, M and Kubicki, M, 'EGATEC - a new high resolution 
engineering model of the global atmospheric electric circuit. I. Currents in the lower 
atmosphere', J. Geophys. Res. (2009), in press.
\item[39.] Palle Bago, E and Butler, C J, 'The influence of cosmic rays on terrestrial 
clouds and global warming', Astronomy and Geophysics, {\bf 41}, (2000) 18.
\item[40.] Parman, S W, `Blood from a stone: water and life on the early Earth', `Water
 on Earth and Beyond', 22, 23 September (2009), Durham University 
water.workshop:durham.ac.uk.
\item[41.] Pudovkin, M I, Vinogradova, N Ya and Veretenenko, S V, 'Variations of 
atmospheric transparency during solar proton events', Geomagn. Aeron., {\bf 37}, (1977)
2, 124. 
\item[42.] Rogers, M J, Sadzinska, M, Szabelski, J, van der Walt, D J and Wolfendale, 
A W, 'A comparison of cosmic ray energy spectra in Galactic spiral arm and interarm 
regions', J. Phys.G: Nucl.,Part. Phys., {\bf 14}, (1988) 1147.
\item[43.] Rycroft, M J, Harrison, R G, Nicoll, K A and Marcev, E A, `An overview of 
Earth's Global Electric Circuit and Atmospheric Conductivity', (2008), Planetary 
Atmospheric Electricity, Eds.F.Leblanc et al., Springer, \\ 
doi:10.1007/978-0-387-87664-1\_6.
\item[44.] Satori, G, Lemperger, I and Bor, J., `Modulation of Annual and Semiannual 
areal variations of global lightning on the 11-y solar cycle' (2007), 2nd Int. Symp. on
 Lightning Physics and Effects, Vienna; European COST action, p.18.
\item[45.] Saunders, R H and Prendergast, K H, 'The possible relation of the 3-kpc arm 
to explosion in the Galactic nucleus', Astrophys. J., {\bf 188}, (1974) 489.
\item[46.] Shaviv , N J, 'Cosmic ray diffusion from the Galactic Spiral Arms, iron 
meteorites and a possible climatic connection', Phys. Rev. Lett., {\bf 89} (2002), 
 051102.
\item[47.] Shaviv, N J, 'The spiral structure of the Milky Way,cosmic rays and ice age 
epochs on Earth', New Astronomy, {\bf 8}, (2003) 39.
\item[48.] Shumilov, O I, Kasatkina, E A, Henriksen, K and Vashenyuk, E, 'Enhancement 
of stratospheric aerosols after solar proton event', Ann. Geophysicae, {\bf 14}, (1996)
 1119.
\item[49.] Sloan, T and Wolfendale, A W, 'Testing the proposed causal link between 
cosmic rays and cloud cover', Environmental Research Letters, {\bf 3}, (2008) 024001.
\item[50.] Starkov, G V and Roldugin, V K, 'On the relation between the variations in 
the atmospheric transparency and geomagnetic activity', Geomagn. Aeron., {\bf 34}, 
(1994) 4, 156.
\item[51.] Stozhkov, Yu I, 'The role of cosmic rays in the atmospheric process',
J. Phys.G: Nucl., Part. Phys. {\bf 29}, (2003) 913.
\item[52.] Svensmark H, 'Cosmoclimatology : a new theory emerges', News Rev. Astron. 
Geophys., \textbf{48}, (2007) 1.18.
\item[53.] Svensmark, H, Bondo, T and Svensmark, J, 'Cosmic ray decreases affect 
atmospheric aerosols and clouds', Geophys. Res. Lett., {\bf 36} (2009) L15101.
\item[54.] Svensmark, H and Friis-Christensen, E, J. 'Variation of cosmic ray flux and 
global cloud coverage: a missing link in sun-climate relationships', Atmos. Solar-Terr.
 Phys., {\bf 59}, (1997) 1225. 
\item[55.] Tinsley, B A, Burns, G B and Zhou, Limin, 'The role of the global electric 
circuit in solar and internal forcing of clouds and climate', Advances in Space 
Research \textbf{40}, (2007) 1126.
\item[56.] Tinsley, B A, 'The global atmospheric circuit and its effect on cloud 
microphysics', Rep. Progr. Phys., {\bf 71}, (2008) 066801.
\item[57.] Tonev, P and Velinov, P I Y, 'Quasi-electrostatic fields in the near-earth 
space produced by lightning and generation of runaway electrons in ionosphere', Adv. 
Space Res., {\bf 31}, (2003) 1443.
\item[58.] Usoskin, I G and Kovaltsov, G A, 'Cosmic ray indused ionization in the 
atmosphere: full modeling and practical applications', J. Geophys. Res., {\bf 111},
(2006) D21206.
\item[59.] Usoskin, I G, Tylka, A J, Kovaltsov, G A and Dietrich W F, 'Long-term 
geomagnetic changes and their possible role in regional atmospheric ionization and 
climate', Proc. 31st ICRC, Lodz (2009) SH3.5-105.
\item[60.] Vahia M N, 'Long term variability of heliopause due to changing conditions 
in local interstellar medium', In: 'Solar Influence on the Heliosphere and Earth's 
environment: Recent progress and prospects', ed. N.Gopalswamy and A.Bhattacharya, ILWS 
and Indian Institute of Geomagnetism, Mumbai, India, (2006), 189.
\item[61.] Velinov, P I Y, Mishev, A and Mateev, L, ' Model for induced ionization by 
galactic cosmic rays in the Earth atmosphere and ionosphere', Adv. Space Res., 
{\bf 44}, (2009) 1002.
\item[62.] Voiculescu, M, Usoskin, I G and Mursula, K, 'Different response of clouds to
 solar input', Geophys. Res. Lett., {\bf 33}, (2006) L21802.
\item[63.] Wdowczyk J and Wolfendale A W, `Cosmic Rays and Ancient Catastrophes', 
Nature, \textbf{268}, (1977) 510.
\item[64.] Williams, E R, 'Encyclopedia of Atmospheric Sciences', ed. J.R.Holton, 
J.A.Pyle, J.A.Curry (Academic Press, New York), (2002) 724.
\item[65.] Wolfendale, A W, 'Cosmic Rays at Ground Level', Ed. A.W.Wolfendale, Inst. of
 Physics, (1973) 1.
\item[66] Yukhimuk, V, Roussel-Dupre, R A, Sympalisty, E M D and Taranenko Y, 'Optical
characteristics of blue jets produced by runaway air breakdown, simulation results', Geophys. Res. Lett., {\bf 25}, (1998) 3289.   
\end{itemize}

\newpage

{\bf Captions to Figures}

{\bf Figure 1}
Short-term variations of cosmic rays over a period of 1 million years using our statistical model (Erlykin and Wolfendale, 2001a).  The 'bin width' is 1000y.  The results relate to a model with an energy dependent diffusion coefficient having exponent $\delta = 0.5$ (in the relation diffusion coefficient D $\alpha$ E$^{\delta}$, where E is the proton energy), and supernova remnants accelerating CR protons up to a maximum energy of 10 PeV.
The intensities at 1 PeV and 10 PeV are displaced upwards by 2 and 10 respectively for ease of discrimination

{\bf Figure 2}
The $`log N, log S'$ plot for peak heights from Figure 1, 68 in all.  Each peak has 
height $`S'$ = the ordinate in Figure 1 minus the previous minimum. The ordinate, 
$N(>S)$, is scaled so that it represents the number of peaks per hundred million years 
of height $>S$.
The lines are simple parabolic fits to the points.  Importantly, in the middle region 
($logS \sim 1$), they have slope $\gamma = 1$, appropriate to a 2-dimensional 
distribution of sources, whereas at high values of $S$ they are as appropriate for a 
3-D distribution of sources

{\bf Figure 3}
Fraction of time (in the My sample) for which the intensity is $\tau$ times larger (in 
logarithmic units) than the overall median value and the fraction for which the 
intensity is $\tau$ times smaller. In terms of evolutionary effects, the excesses and deficits are both of importance.

{\bf Figure 4}
As for Figure 1 but for a bin width of 10,000 y

{\bf Figure 5}
The CR intensity from a single supernova at 300 pc from the Earth. Our estimated range 
of the `present time' is indicated by vertical dotted lines.

\newpage
\begin{figure}
\begin{center}
\epsfxsize=15cm\epsfbox{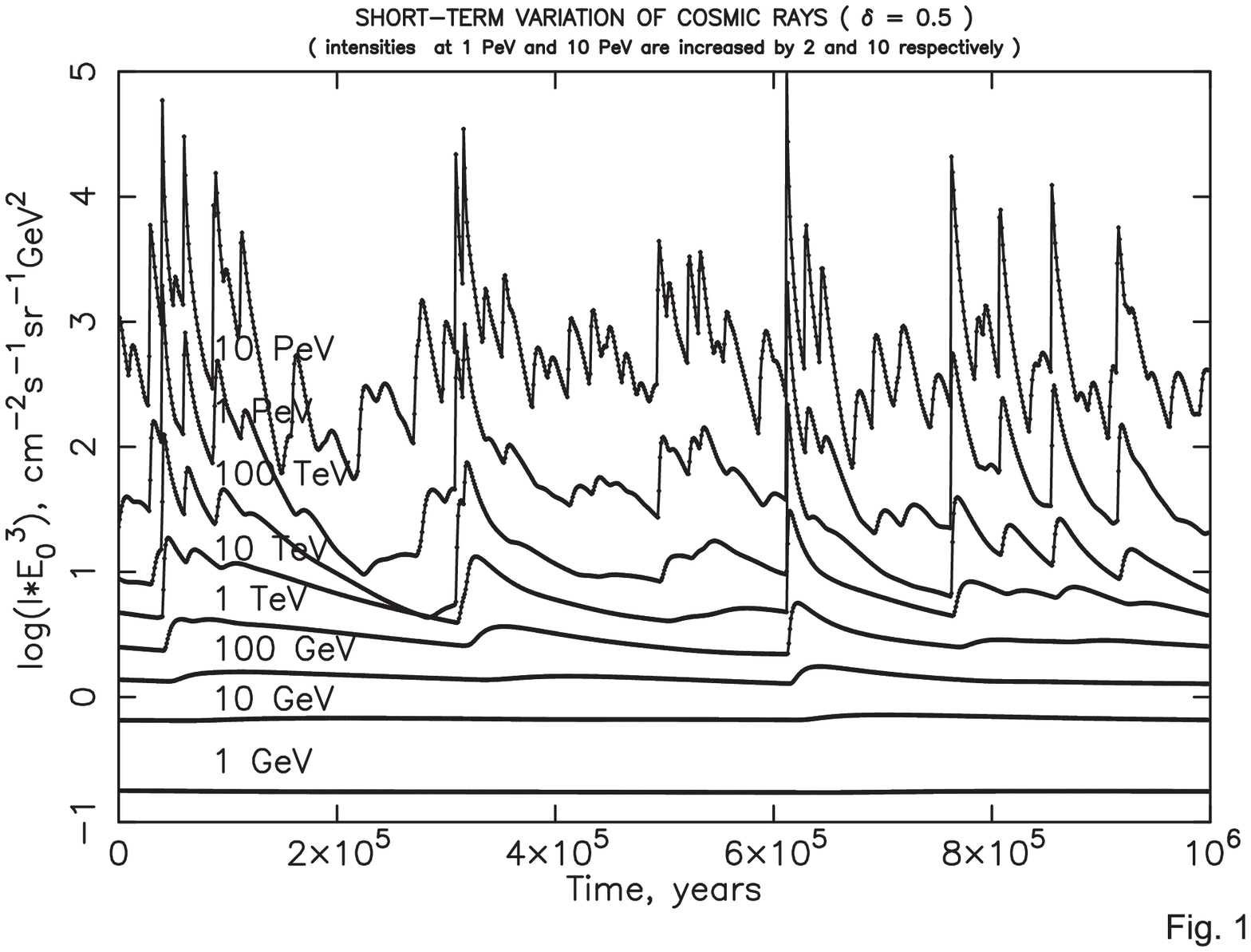}
\end{center}
\end{figure}

\newpage
\begin{figure}[hptb]
\begin{center}
\includegraphics[height=15cm,width=12cm,angle=-90]{lightning_fig2_R2.eps}
\end{center}
\end{figure}

\newpage
\begin{figure}
\begin{center}
\epsfxsize=15cm\epsfbox{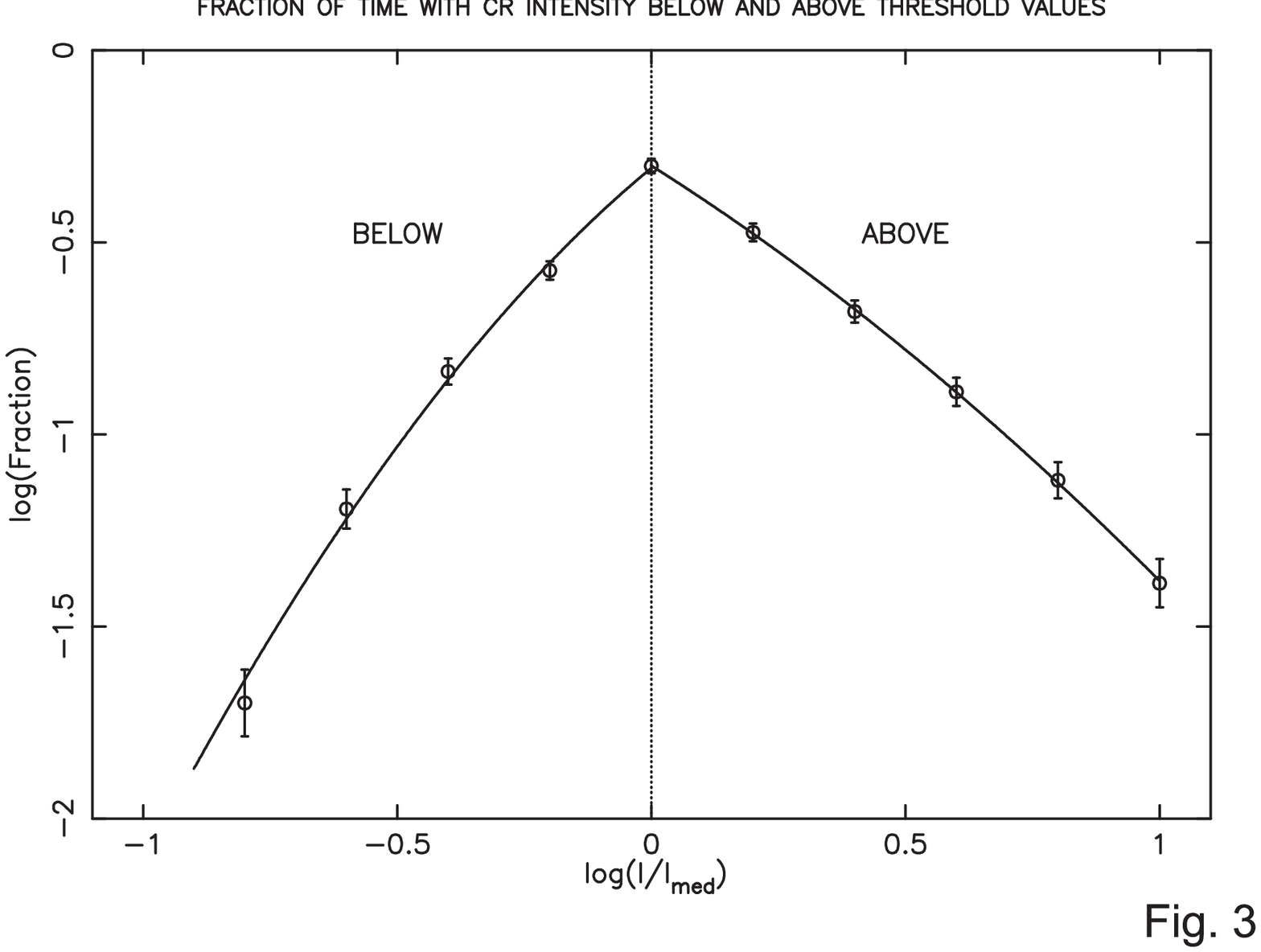}
\end{center}
\end{figure}

\newpage
\begin{figure}
\begin{center}
\epsfxsize=15cm\epsfbox{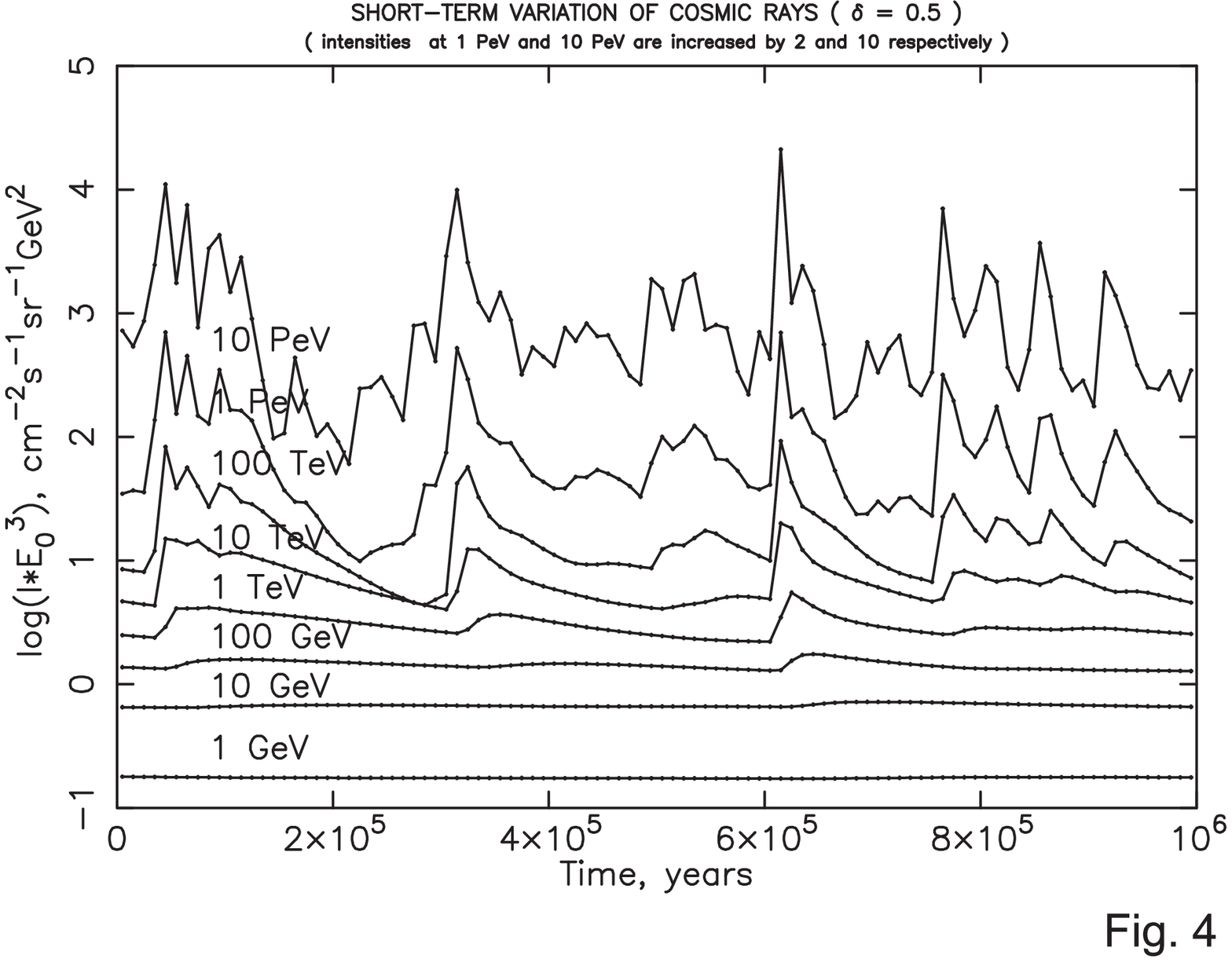}
\end{center}
\end{figure}

\newpage
\begin{figure}
\begin{center}
\epsfxsize=10cm\epsfbox{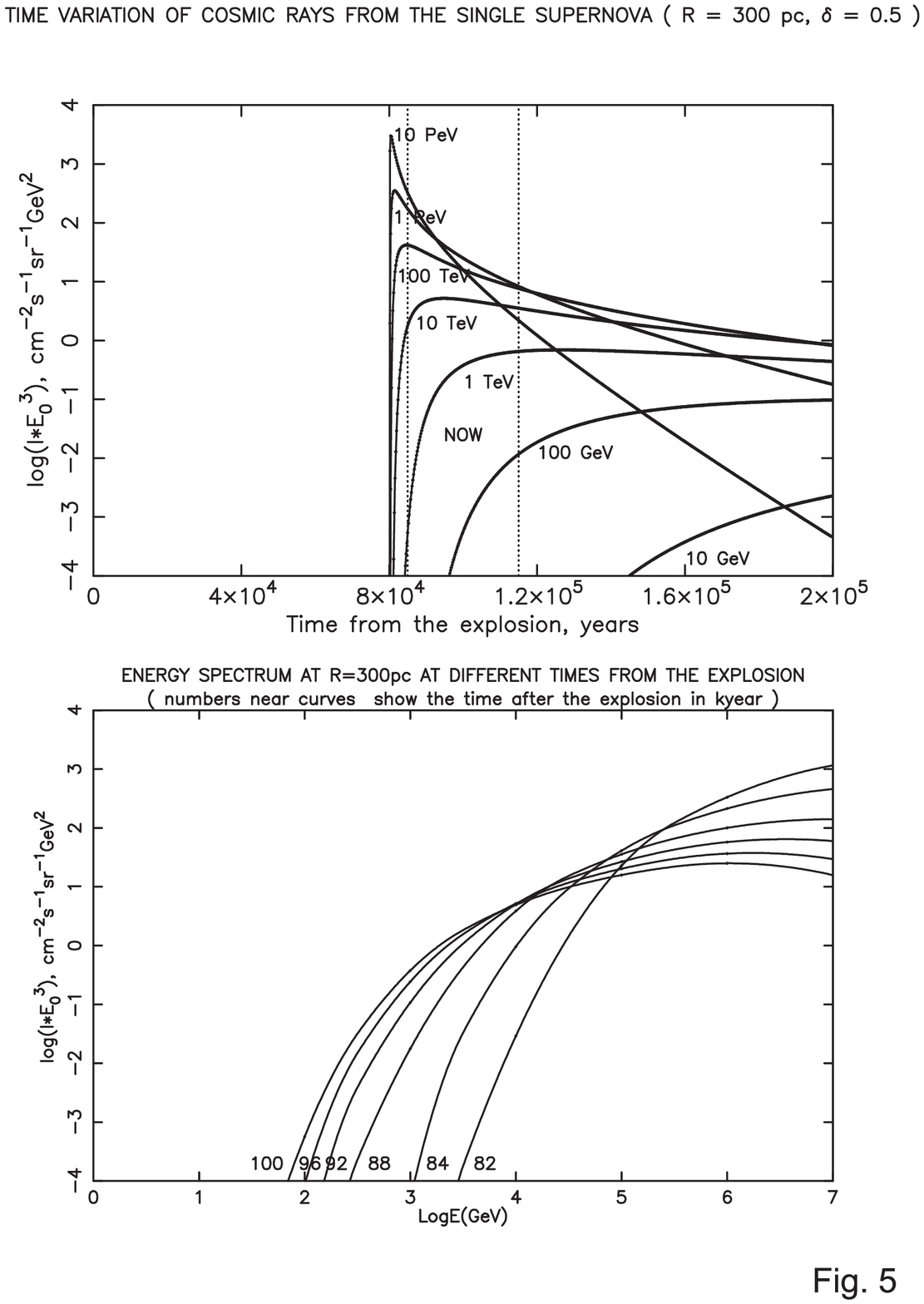}
\end{center}
\end{figure}

\end{document}